\documentclass[12pt]{article}
\usepackage{pdproc,epsfig}
\usepackage{graphics}
\usepackage{graphicx}

  \def\lsim{\buildrel < \over {_{\sim}}}

\begin{document}
\begin{flushright}
  FERMILAB-CONF-09-196-T\\
\end{flushright}
\vspace*{0.1in}
\renewcommand{\thefootnote}{\alph{footnote}}

\title{OVERVIEW OF NEUTRINO MIXING MODELS AND 
   WAYS TO DIFFERENTIATE AMONG THEM}

\author{ CARL H. ALBRIGHT}

\address{Department of Physics, Northern Illinois University,
  DeKalb, IL 60115, USA\\}
  
  \centerline{\footnotesize and}
  
\address{Theoretical Physics Department,
  Fermi National Accelerator Laboratory,
  Batavia, IL 60510, USA\\
  {\rm E-mail: albright@fnal.gov}} 

\abstract{An overview of neutrino-mixing models is presented 
  with emphasis on the types of horizontal flavor and vertical 
  family symmetries that have been invoked. Distributions for
  the mixing angles of many models are displayed.
  Ways to differentiate among the models and to narrow the 
  list of viable models are discussed.}

\normalsize\baselineskip=15pt

\section{Introduction}
  Several hundred models of neutrino masses and mixings
  can be found in the literature which purport to explain the 
  known oscillation data and predict the currently unknown
  quantities.  We present an overview of the types of models
  proposed and discuss ways in which the list of viable 
  models can be reduced when more precise data is obtained. 
  This presentation is an update of one published in 2006 in
  collaboration with Mu-Chun Chen\cite{a-c}.	

\section{Present Oscillation Data and Unknowns}
  The present data within $3\sigma$ accuracy as determined by 
  Fogli et al.\cite{fogli}, for example, is given be 
  \begin{eqnarray}
   \label{eq:data} 
    \Delta m^2_{32} &=& 2.39\begin{array}{c} +0.42 \\ -0.33 \\ \end{array}
    \times 10^{-3}\ {\rm eV^2,} \nonumber \\
     \Delta m^2_{21} &=& 7.67\begin{array}{c} +0.52\\ -0.53\\ \end{array}
    \times 10^{-5}\ {\rm eV^2}, \nonumber \\
    \sin^2 \theta_{23} &=& 0.466\begin{array}{c} +0.178\\ -0.135\\ \end{array},
      \nonumber \\
    \sin^2 \theta_{12} &=& 0.312\begin{array}{c} +0.063 \\ -0.049\\ 
      \end{array}, \nonumber \\
    \sin^2 \theta_{13} &\leq& 0.046, \quad (0.016 \pm 0.010), 
   \end{eqnarray}

\noindent  where the last figure in parenthesis indicates a departure 
of the reactor neutrino angle from zero with one $\sigma$
accuracy determination.
The data suggests the approximate tri-bimaximal 
mixing texture of Harrison, Perkins, and Scott\cite{tbm},
\begin{equation}
  U_{PMNS} = \left( \begin{array}{ccc} 2/\sqrt{6} & 1/\sqrt{3} & 0\\
             -1/\sqrt{6} & 1/\sqrt{3} & -1/\sqrt{2}\\
	   -1/\sqrt{6} & 1/\sqrt{3} & 1/\sqrt{2} \end{array}\right),
\label{eq:PMNS}
\end{equation}

\noindent with $\sin^2 \theta_{23} = 0.5,\ \sin^2 \theta_{12} = 0.33$,
and $\sin^2 \theta_{13} = 0$.

The reason for the plethera of models still in agreement with experiment
of course can be traced to the inaccuracy of the present data and the 
imprecision of the model predictions in many cases.  In 
addition, there are a number of unknowns that must still be determined:  
the hierarchy and absolute mass scales of the light neutrinos; the Dirac 
or Majorana nature of the neutrinos; the CP-violating phases of the mixing 
matrix; how close to zero the reactor neutrino angle, $\theta_{13}$, lies;
how near maximal the atmospheric neutrino mixing angle is; whether the 
approximate tri-bimaximal mixing is a softly-broken or an accidental symmmetry;
whether neutrino-less double beta decay will be observable, and how large 
charged lepton flavor violation will turn out to be.
In this presentation we survey the models to determine what they predict
for the mixing angles, hierarchy, and briefly mention the role charged 
lepton flavor violation can play.
    
\section{Theoretical Framework}
The observation of neutrino oscillations implies that neutrinos have mass,
with the mass squared differences given in Eq.(\ref{eq:data}).  Information
concerning the absolute neutrino mass scale has been determined by the 
combined WMAP, SDSS, and Lyman alpha data which place an upper limit on 
the sum of the masses\cite{masssum},
\begin{equation}
  \sum_i m_i \leq 0.17 - 1.2\ {\rm eV},
\label{eq:masssum}
\end{equation}

\noindent depending upon the conservative nature of the bound extracted.
An extension of the SM is then required, and possible approaches 
include one or more of the following:
\begin{itemize}
\item the introduction of dim-5 effective non-renormalizable operators;
\item the addition of right-handed neutrinos with their Yukawa couplings
  to the left-handed neutrinos;
\item the addition of direct mass terms with right-handed Majorana couplings;
\item the addition of a Higgs triplet with left-handed Majorana couplings;
\item the addition of a fermion triplet with Higgs doublet couplings.
\end{itemize}

If we exclude the last possibility, the general $6 \times 6$ neutrino mass 
matrix in the $B(\nu_{\alpha L},\ N^c_{\alpha L})$ flavor basis of the six 
left-handed fields then has the following structure in terms of $3 \times 3$ 
submatrices:
\begin{equation}
  \cal{M} = \left(\begin{array}{cc} M_L & M^T_N \\ M_N & M_R \end{array}
             \right),
\label{eq:6x6}
\end{equation}

\noindent where $M_N$ is the Dirac neutrino mass matrix, $M_L$ the 
left-handed and $M_R$ the right-handed Majorana neutrino mass matrices.
With $M_L = 0$ and $M_N << M_R$ the type~I seesaw formula,  
\begin{equation}
  m_\nu = - M^T_N M^{-1}_R M_N,
\label{eq:typeI}
\end{equation}

\noindent is obtained for the light left-handed Majorana neutrinos, while if 
$M_L \neq 0$ and $M_N << M_R$, one obtains the type II seesaw formula,
\begin{equation}
  m_\nu = M_L - M^T_N M^{-1}_R M_N.
\label{eq:typeII}
\end{equation}

There are two main approaches which we now describe that one can 
pursue to learn more about the theory behind the lepton mass generation.

\subsection{Top - Down Approach}
In the top-down approach one postulates the form of the mass matrix from
first principles.  The models will differ then due to the horizontal flavor 
symmetry chosen, the vertical family symmetry (if any) selected, and the 
fermion and Higgs representation assignments made.

The effective light left-handed Majorana mass matrix $m_\nu$ is constructed 
directly or with the seesaw formula once the Dirac neutrino matrix $M_N$ 
and the Majorana neutrino matrices $M_R$ (and $M_L$) are specified.
Since $m_\nu$ is complex symmetric, it can be diagonalized by a unitary 
transformation, $U_{\nu_L}$, to give
\begin{equation}
  m^{diag}_\nu = U^T_{\nu_L} m_\nu U_{\nu_L} = {\rm diag}(m_1,\ m_2,\ m_3),
\end{equation}

\noindent with real, positive masses down the diagonal.  On the other hand,
the Dirac charged lepton mass matrix is diagonalized by a bi-unitary 
transformation according to 
\begin{equation}
  m^{diag}_\ell = U^\dagger_{\ell R} m_\ell U_{\ell L} = 
  {\rm diag}(m_e,\ m_\mu, m_\tau).
\end{equation}

\noindent The neutrino mixing matrix\cite{PMNS}, $V_{PMNS}$, is then given 
by \\
\begin{equation}
  V_{PMNS} \equiv U^\dagger _{\ell L} U_{\nu_L} = U_{PMNS}\Phi,\\
\label{eq:VPMNS}
\end{equation}

\noindent   in the lepton flavor basis with $\Phi = {\rm diag}(1,e^{i\alpha},
\ e^{i\beta})$.  Note that the Majorana phase matrix 
$\Phi$ is required in order to compensate for any phase rotation on 
$U_{\nu_L}$ needed to bring it into the Particle Data Book phase 
convention\cite{pdb}.
  
\subsection{Bottom - Up Approach}
On the other hand, with a bottom-up approach in the diagonal lepton flavor
basis and with the general PMNS mixing matrix, one can determine 
the general texture of the light neutrino mass matrix to be 
\begin{eqnarray}
\label{eq:bot-up}
 M_\nu &=& U^*_{PMNS}\Phi^* M^{\rm diag}_\nu \Phi^*U^\dagger_{PMNS}\nonumber\\
	&=& U^*_{PMNS}{\rm diag}(m_1,
\ m_2 e^{-2i\alpha},\ m_3 e^{-2i\beta})U^\dagger_{PMNS}\nonumber\\[0.1in]
&\equiv& \left(\begin{array}{ccc}
A & B & B'\\ \cdot & F' & E\\ \cdot & \cdot & F\\ \end{array}\right),
\end{eqnarray}

\noindent  where the matrix elements are expressed in terms of the unknown
neutrino masses, mixing angles and phases.  By restricting the mixing matrix,
one can learn that some of the matrix elements may not be independent.
  
\section{Models and Mixing Angle Predictions}
When the first hints of atmospheric neutrino oscillations were 
discovered around 1992 by the Super-Kamiokande Collaboration\cite{s-k}, it 
became fashionable to assign texture zeros in different positions to 
$m_\nu$ with a top-down approach in hopes of identifying some flavor 
symmetry, but the procedure is basis dependent\cite{textzero}. 

Another popular method invoked a $L_e - L_\mu - L_\tau$ lepton flavor
symmetry\cite{e-mu-tau}.  The mass matrix then assumes the following form
\begin{equation}
  m_\nu = \left(\begin{array}{ccc} 0 & * & * \\ \cdot & 0 & 0 \\ 
    \cdot & 0 & 0 \\ 
  \end{array}
  \right), \\
\end{equation}

\noindent which only leads to an inverted hierarchy.

By making use of a bottom-up approach instead, one is able to observe that a 
$\mu~-~\tau$ interchange symmetry with $B' = B,\ F' = F$ in 
Eq. (\ref{eq:bot-up}) leads to $\sin^2 \theta_{23} = 0.5,\ \sin^2 \theta_{13} 
= 0$ with $\sin^2 \theta_{12}$ arbitrary.  

On the other hand, with the assumption of exact tri-bimaximal mixing for
which $\sin^2 \theta_{23} = 0.5,\ \sin^2 \theta_{13} = 0$, and $\sin^2 
  \theta_{12} = 0.333$, one finds in Eq. (\ref{eq:bot-up}) that $B' = B,
  \ F' = F = \frac{1}{2}(A + B + D)$ and $E = \frac{1}{2}(A + B - D)$, so 
that just three unknowns are present.

With the realization in the past five years that neutrino mixing is well 
approximated by the tri-bimaximal mixing matrix, the name of the game 
has become one of finding what discrete horizontal flavor symmetry groups 
would lead naturally to this mixing pattern.  Such flavor symmetries can 
then be used as starting points with soft breaking as the next approximation.
  
\subsection{Discrete Horizontal Flavor Symmetry Groups}
Of special interest are those groups containing doublet and triplet
irreducible representations.   We list several of the well-studied groups
and pertinent features of each.

The permutation group of three objects, $S_3$, contains 6 elements with 
$1, 1',$ and 2 dimensional irreducible representations (IR's).  The same 
eigenstates occur as those for tri-bimaximal mixing, but there is a 2-fold 
neutrino mass degeneracy.

The group $A_4$ of even permutations of four objects has 12 
elements with IR's labeled $1, 1', 1''$, and 3.  A $U(1)_{FN}$ flavon 
group\cite{f-n} 
is often imposed to fix the mass scale which is otherwise scale-independent.
Early attempts to extend this flavor group to the quark sector failed, as 
the CKM mixing matrix for the quarks remained diagonal.

The group $T'$ is the covering group of $A_4$, but interestingly $A_4$ 
is not one of its subgroups.  It contains 24 elements with $1, 1', 1'', 3,
2, 2', 2''$ IR's, where the first four are identical to those in $A_4$.   
While tri-bimaximal mixing is obtained for the leptons, due to the presence 
of the three doublet IR's, a satisfactory CKM mixing matrix can also be 
obtained for the quarks.

The permutation group of 4 objects, $S_4$, has 24 elements with 
$1, 1',2, 3, 3'$ IR's.  Lam has proved this is the smallest symmetry 
group naturally related to tri-bimaximal mixing, if one requires all IR's to 
participate in the model\cite{lam}.  

\subsection{Examples Involving GUT Models}
Studies of neutrino mixing models in the framework of grand unified 
theories with a vertical family symmetry were first pursued in the 1990's 
and more intensely following the discovery of atmospheric neutrino 
oscillations by the Super-Kamiokande Collaboration in 1998.  Examples exist
of models based on $SU(5)$, $SO(10)$, and $E_6$, where the 
$SO(10)$ models are generally of two types.

The so-called ``minimal'' $SO(10)$ models\cite{minimal} involve Higgs fields 
appearing in the ${\bf 10}$ and ${\bf 126}$ IR's, but newer models of this 
type have been extended to include the ${\bf 120}$, ${\bf 45}$, and/or 
${\bf 54}$ IR's.
They generally result in symmetric and/or antisymmetric contributions
to the quark and lepton mass matrices.

On the other hand, $SO(10)$ models\cite{lopsided} with Higgs fields in the 
${\bf 10, 16, \overline{16}}$ and ${\bf 45}$ IR's result in ``lopsided''
down quark and charged lepton mass matrices due to the $SU(5)$
structure of the electroweak VEV's appearing in the ${\bf 16}$ and 
$\overline{\bf 16}$ representations.

For either type of GUT model, type I seesaws only lead to a stable
normal hierarchy for the light neutrino masses\cite{alb}, while type I + II 
seesaws can also result in an inverted hierarchy.  Most of the 
$SO(10)$ models have a continuous and/or discrete flavor 
symmetry group producted with them, but no efforts were initially made to 
introduce a discrete flavor symmetry group of the type discussed
earlier.  A few examples can now be found in the literature which 
combine an $SU(5),\ SO(10)$ or $E_6$ GUT symmetry with 
a $T'$ or $A_4$ flavor symmetry with some success\cite{famflav}.

\begin{table}[h!]
\caption{Mixing Angles for Models with Lepton Flavor Symmetry.}\label{tab:LFS}
\vspace*{0.1in}
\small
\begin{tabular}{||ll|c|c|c|c|c||}\hline\hline
{} &{} &{} &{} &{} &{} &{}\\[-0.05in]
\multicolumn{2}{||l|}{Reference} & {Hierarchy}  & 
  {\ $\sin^2 2\theta_{23}$} & {\ $\sin^2 \theta_{12}$}  & 
  {$\sin^2 \theta_{13}$}  & {\ $\sin^2 \theta_{23}$}\\
{} &{} &{} &{} &{} &{} &{}\\[-0.05in]
\hline
\multicolumn{3}{||l|}{\bf Texture Zero Models:} &   &    &  &  \\
GL1 & \cite{GL1} & NH  & 1.0  &       & $\geq$ 0.005     & \\
WY  & \cite{WY}  & NH  &	   &       & 0.0006 - 0.0030  & \\
    &            & IH &	   &       & 0.0006 - 0.0030  & \\
    &            & NH  &	   &       & $< 0.023$        & \\
    &            & NH  &	   &       & 0.017 - 0.14     & \\
CPP & \cite{CPP}    & NH  & 	   &       & 0.0066 - 0.0083  & \\
       &         & IH &	   &       & $\geq 0.00005$   & \\
       &         & IH &	   &       & $\geq 0.032$     & \\
\hline 
\multicolumn{3}{||l|}{\bf $\mathbf{ L_e - L_\mu - L_\tau}$ Models:} &  & & & \\
BM   & \cite{BM}   & IH &      &               & 0.00029      & \\
GMN1 & \cite{GMN1}  & IH &      & $\geq 0.28$   & $\leq 0.05$  & \\
PR   & \cite{PR}   & IH &      & $\lsim 0.37$  & $\geq 0.007$ & \\
GL2  & \cite{GL2}  & IH &      & 0.30          & 0            & \\
\hline
\multicolumn{3}{||l|}{\bf 2-3 Symmetric Models:}  &  &  &  & \\
RS    & \cite{RS}   & NH    & $\theta_{23} \leq 45^\circ$ &  & 0    & \\
      &             & IH   & $\theta_{23} \geq 45^\circ$ &  & $\leq 0.02$
   & \\
MN    & \cite{MN}   & NH    & 1.0    &       & 0.0024     & \\
AKKL  & \cite{AKKL} & NH    &	  &       & 0.006 - 0.016 & \\
      &             & IH   &        &       & 0.022 - 0.04  & \\
SRB   & \cite{SRB}  & IH   & 1.0    & 0.31  & 0         & 0.50 \\
BY    & \cite{BY}   & NH    & 1.0    & 0.33  & $< 0.0025$ & \\
      &             & IH   & 1.0    & 0.33  & $< 0.008 $ & \\
\hline
\multicolumn{2}{||l|}{\bf $\mathbf{S_3}$ Models:} &  &  &  &  &   \\
KMMR-J & \cite{KMMR-J} & IH   & 1.0    &      & 0.000012     & \\
CFM    & \cite{CFM}    & NH    &	  &         & 0.00006 - 0.001 & \\
T      & \cite{T}      & NH    & 	  &         & 0.0016 - 0.0036 & \\
TY     & \cite{TY}     & IH   & 0.93   & 0.30   & 0.0025      & 0.37  \\
MNY    & \cite{MNY}    & NH    &	  &         & 0.000004 - 0.000036 & \\
MMP    & \cite{MMP}    & IH   & 1.0    & 0.31 & 0.0034          & \\
MC     & \cite{MC}     & NH    & 1.0    &      & $< 0.01$        & \\
\hline
\end{tabular}
\end{table}
\newpage
\begin{table}
\small
\begin{tabular}{||ll|c|c|c|c|c||}\hline
{} &{} &{} &{} &{} &{} &{}\\[-0.1in]
\multicolumn{2}{||l|}{Reference} & {Hierarchy}  & 
  {\ $\sin^2 2\theta_{23}$} & {\ $\sin^2 \theta_{12}$}  & 
  {$\sin^2 \theta_{13}$}  & {\ $\sin^2 \theta_{23}$}\\
{} &{} &{} &{} &{} &{} &{}\\[-0.1in]
\hline
\multicolumn{3}{||l|}{\bf $\mathbf{A_4}$ Tetrahedral Models:} & & & &  \\
Ma1    & \cite{Ma1}     & NH    & 1.0    &  0.31         & 0       & 0.50 \\
       &             & IH   & 1.0  &  0.33 - 0.34  & 0       & 0.50 \\
ABGMP & \cite{ABGMP} & NH & 1.0  &\  0.27 - 0.30\ & 0.0007 - 0.0037 & 
  0.51 - 0.52
  \\
AG1   & \cite{AG1}    & NH & 1.0 & 0.31 & 0.0026 - 0.034 & 0.51 - 0.56 \\
HT    & \cite{HT}    & NH    & 1.0    & 0.29 - 0.33   & $< 0.0022$ &   \\
AG2   & \cite{AG2}   & IH   & 1.0    & 0.27 - 0.34   & $< 0.0012$ & 
  0.52 - 0.53 \\
L     & \cite{L}     & NH    & 1.0    & 0.29 - 0.38   & 0.0025  &   \\
Ma2   & \cite{Ma2}    & NH    & 1.0    & 0.32          & 0     &  0.50 \\ 
\hline
\multicolumn{2}{||l|}{\bf $\mathbf{S_4}$ Models:}  &  &  &  &  &  \\
MPR   & \cite{MPR}   & Q-deg  & 0.99   & 0.25 - 0.37  & 0.008 - 0.01 & 
  0.44\\
HLM   & \cite{HLM}   & NH    & 1.0       & 0.30     & 0.0044  & 0.50  \\
      &              & NH    & 1.0       & 0.31     & 0.0034  & 0.50  \\
Z     & \cite{Z}     & NH    & 0.96 - 1.0  & 0.311  & $< 0.030$ & 
  0.41 - 0.50 \\
\hline
\multicolumn{2}{||l|}{\bf $\mathbf{SO(3)}$ Models:}  &  &  &  &  &  \\
M    & \cite{M}    & NH	 & 1.0       & 0.31     & 0.00005 &  \\
W    & \cite{W}     & NH    &           &          & 0.0027 - 0.036 &  \\
\hline
\multicolumn{2}{||l|}{\bf $\mathbf{T'}$ Models:}  &  &  &  &  &  \\
FM    & \cite{FM}    & NH	 & 0.93 - 0.95 &        & 0.024 - 0.036 &  \\
\hline\hline
\end{tabular}
\end{table}

\begin{table}[h]
\caption{Mixing Angles for Models with Sequential Right-Handed 
  Neutrino Dominance.}\label{tab:SRND}
\vspace*{0.1in}
\small
\begin{tabular}{||ll|c|c|c|c|c|c||}\hline\hline
{} &{} &{} &{} &{} &{} &{} &{} \\[-0.1in]
\multicolumn{2}{||l|}{Reference} & Flavor Sym. & {Hierarchy}  & 
  {\ $\sin^2 2\theta_{23}$} & {\ $\sin^2 \theta_{12}$}  & 
  {$\sin^2 \theta_{13}$}  & {\ $\sin^2 \theta_{23}$}\\
{} &{} &{} &{} &{} &{} &{} &{} \\[-0.1in]
\hline
D    & \cite{D}	& Z$_3$ & NH	& 	&	& 0.008 - 0.14 &  \\
K    & \cite{K} & SO(3) & NH	& 0.99 - 1.0	& 0.28 - 0.39 & 0.0027 & \\
H    & \cite{H} &   &  NH       & 1.0	& 0.30	& 0.0033 & 0.52 \\
EH   & \cite{EH} & U(1) & NH	& 0.99	& 0.31	& 0.0009 & 0.54 \\
\hline\hline 
\end{tabular}
\end{table}

\section{Survey of Mixing Angle Predictions}
The author has updated a previous survey\cite{a-c} made in collaboration
with Mu-Chun Chen in 2006 of models in the literature which satisfied
the then current experimental bounds on the mixing angles and gave
reasonably restrictive predictions for the reactor neutrino angle.
The cutoff date for the present update is January 2009.

Many models in the literature lack firm predictions for any of the mixing 
angles.  For our analyzis no requirement is 
made that the solar and atmospheric mixing angles or the mass differences
be predicted, but if so, they must also satisfy the bounds given in 
Eq. (\ref{eq:data}).  The 86 models which meet our criteria are listed in
Tables 1 - 4.

\begin{table}
\caption{Mixing Angles for $SO(10)$ Models 
  with Symmetric/Antisymmetric Contributions.}\label{tab:so10sym} 
\vspace*{0.1in}
\small
\begin{tabular}{||ll|c|c|c|c|c|c||}\hline\hline
{} &{} &{} &{} &{} &{} &{} &{} \\[-0.1in]
\multicolumn{2}{||l|}{Reference} & Flavor Sym. &Hier.& 
  {\ $\sin^2 2\theta_{23}$} & {\ $\sin^2 \theta_{12}$}  & 
  {$\sin^2 \theta_{13}$}  & {\ $\sin^2 \theta_{23}$}\\
{} &{} &{} &{} &{} &{} &{} &{}  \\[-0.1in]
\hline 
BRT  &\cite{BRT}& U(2) x U(1)$^n$ & NH & 0.99  & 0.26  & 0.0024 & 0.55  \\
BW   &\cite{BW}&       & NH   &	   &	        & O(0.01) & \\
SP   &\cite{SP}&        & NH   & 0.99 & 0.30 	& 0.0002 & 0.50 \\
Ra   &\cite{Ra}& SU(2) x U(1) & NH & 	&	& O(0.01) &  \\
BO   &\cite{BO}& U(1)$_A$ & NH & $\geq 0.95$ &\ 0.19 - 0.38\ & 0.0014 & \\
O    &\cite{O}&      & NH	& 0.94	& 0.31 	& 0.0007 &  \\
KR   &\cite{KR}& SU(3) x R &  NH  & 0.93 & 0.30 & 0.058 & 0.63 \\
     &           & x U(1) x Z$_2$ &  &  &  & & \\  
Ro   &\cite{Ro}&        & NH   &	&	& 0.0056 &   \\
     &         &        & IH  &	&	& 0.036  &   \\
GMN2 &\cite{GMN2}&        & NH	& $\leq 0.91$ & $\geq 0.34$ & 0.026 &  \\
YW   &\cite{YW}& 	& NH   & 0.96	& 0.29	& 0.04 &  \\
CM1  &\cite{CM1}& SU(2) x Z$_2$ & NH  & 1.0  & 0.26 & 0.014 & 0.51 \\
     &         &  x Z$_2$ x Z$_2$ &  &  &  & & \\
BeM  &\cite{BeM}& 	& NH   & 0.93  & 0.29	& 0.012  & 0.53 \\
BaM  &\cite{BaM}&	& NH   & 0.98  & 0.31 	& 0.013  & \\
DR   &\cite{DR}& D$_3$ x U(1)   & NH & 0.99 & 0.29 & 0.0024 & 0.55 \\
     &         & x Z$_2$ x Z$_2$ &   &  &  & & \\
VR   &\cite{VR}& SU(3)  & NH   & $\geq 0.99$ & 0.29 - 0.38 & 0.024 & 
  0.44 - 0.56 \\
DMM  &\cite{DMM}& 	& NH	& 	&	& 0.0036 -  &  \\
     &          &       &       &       &       & 0.012     &  \\     
ShT   &\cite{ShT}& R sym. & NH   & 0.99 & 0.31	& 0.0001 -  & 0.44  \\
     &          &       &       &       &       & 0.04     &  \\     
BN   &\cite{BN}& SU(3)  & NH   & 1.0   & 0.26-0.28  & 0.0009 - &  0.5 - 0.51 \\
     &          &       &       &       &       & 0.016     &  \\     
BMSV &\cite{BMSV}&  	& IH & 	   &    	& $\geq 0.01$ &  \\
DHR  &\cite{DHR}& D$_3$  & NH   & 1.0  & 0.29  & 0.0025 - &   0.53 - 0.54 \\
     &          &       &       &       &       & 0.0037     &  \\     
KM   &\cite{KM}& SO(3) 5D & NH &         & 0.30 - 0.37  & 0.0012 & \\
CY   &\cite{CY}& S$_4$  & NH   & 1.0        & 0.28  & 0.0029 & 0.53 \\
GK1  &\cite{GK1}& Z$_2$  & NH   &         & 0.031  & 0.01 & \\
FMN  &\cite{FMN}&        & NH   & 1.0        & 0.32  & 0.0002 & 0.53 \\
GK2  &\cite{GK2}& A$_4$ & NH  & $\geq 0.96$ & 0.25 - 0.5 & 0.0002 & 0.4 - 0.7 
  \\
     &           &       & IH &       & 0.28 - 0.5 & 0.0025 & 0.3 - 0.7\\
Mo   &\cite{Mo}&       & NH  & 0.97       & 0.35       & 0.017 & 0.42 \\
P    &\cite{P}& S$_4$ & NH  & 1.0    & 0.26 - 0.38 & 0.0027 - & 0.52 - 0.54 \\
     &          &       &       &       &       & 0.0032     &  \\     
BR   &\cite{BR}&       & NH  &            &            & 0.0027 - & \\
     &          &       &       &       &       & 0.024     &  \\     
\hline\hline
\end{tabular}
\end{table}

\begin{table}[t]
\caption{Mixing Angles for $SO(10)$ Models (or otherwise indicated) with 
  Lopsided Mass Matrices.}\label{tab:soLop} 
\vspace*{0.1in}
\small
\begin{tabular}{||ll|c|c|c|c|c|c||}\hline\hline
{} &{} &{} &{} &{} &{} &{} &{} \\[-0.1in]
\multicolumn{2}{||l|}{Reference} & Flavor Sym. & Hier.  & 
  {\ $\sin^2 2\theta_{23}$} & {\ $\sin^2 \theta_{12}$}  & 
  {$\sin^2 \theta_{13}$}  & {\ $\sin^2 \theta_{23}$}\\
{} &{} &{} &{} &{} &{} &{} &{}  \\[-0.1in]
\hline 
Mae   & \cite{Mae}   & U(1) & NH  &         &	  & 0.048 & \\
AB    & \cite{AB}   & U(1) x (Z$_2)^2$ & NH  & 0.99  & 0.33  & 0.0002 & 0.54\\
BB    & \cite{BB}   &  &  NH	 & 0.97   & 0.29   & 0.0016 - 0.0025 & 0.58 \\
A     & \cite{A}    & U(1) x (Z$_2)^2$ & NH & 0.99  & 0.28   & 0.0022 & 0.55 \\
JLM   & \cite{JLM}  &   &  NH	  & 1.0	  & 0.29	 & 0.019 & 0.49 \\
$^*$ CM2 & \cite{CM2}  & T$'$ x (Z$_2)^2$ & NH  & 1.0  & 0.30  & 0.0030 & 0.50 
  \\
$^\dagger$ StT & \cite{StT} & SU(3) x Z$_2$ & IH & 1.00  & 0.31 & 0.012 & 0.47 
  \\
FHLR  & \cite{FHLR}  &  &  NH   &       &     & 0.05  & \\
      &            &  &  IH &       &     & $\lsim 0.01$ &  \\
HSS   & \cite{HSS} & $\Sigma$(81) & NH   & 1.0   & 0.27   & 0.0004 & 0.53 \\
\hline\hline
\multicolumn{3}{l}{$^*\ SU(5)$ based model}\\
\multicolumn{3}{l}{$^\dagger\ E_6$ based model}\\[-0.05in]
\end{tabular}
\end{table}

Histograms are plotted against 
$\sin^2 \theta_{13}$, where all models are given the same area,
even if they extend across several basic intervals.  
The results are shown in Figs. 1 and 2 for the lepton flavor 
models and grand unified models, respectively.  Two thirds of
both types of models predict $0.001 \lsim \sin^2 \theta_{13} \lsim 0.05$,
while the lepton flavor models have a much longer tail extending
to very small reactor neutrino angles.  The planned  
experiments involving Double CHOOZ and Daya Bay reactors\cite{reac}
will reach down to $\sin^2 2\theta_{13} \lsim 0.01$, so roughly 
two-thirds of the models will be eliminated if no $\bar{\nu}_e$ 
depletion is observed.  Both the T2K Collaboration at JPARC and 
the NO$\nu$A Collaboration at Fermilab are also expected to probe
a similar reach with their $\nu_\mu$ neutrino beams\cite{T2KNOVA}.

Even if $\bar{\nu}_e$ depletion is observed with some accuracy,
it is apparent from the two histograms that the order of 10 - 20
models may survive which must still be differentiated.  One 
suggestion is to make scatterplots of $\sin^2 \theta_{13}\ vs.\ 
\sin^2 \theta_{12}$ and $\sin^2 \theta_{12}\ vs.\ 
\sin^2 \theta_{23}$.  We have attempted to do this in Figs.
3, 4, and 5 for both the lepton flavor models and grand unified
models.  Note that even fewer of the 86 models considered make 
predictions for the solar and atmospheric neutrino mixing angles.  
In addition, we emphasize that only the central value predictions have 
been plotted, while some of the models have rather large
theoretical error bars associated with them.  

Still one can make some interesting conclusions.  In particular,
most of the models considered favor central values of 
$\sin^2 \theta_{12}$ lying below 0.333, the value for exact
tri-bimaximal mixing.  This is in agreement with the present
value extracted in Eq. (\ref{eq:data}).  But perhaps even more
surprising is that central values for $\sin^2 \theta_{23} \geq 0.5$ are
preferred, while the best extracted value of 0.466 from 
Eq. (\ref{eq:data}) lies below 0.5.

\begin{figure}
\begin{center}
\vspace*{-0.3in]}
\mbox{\epsfig{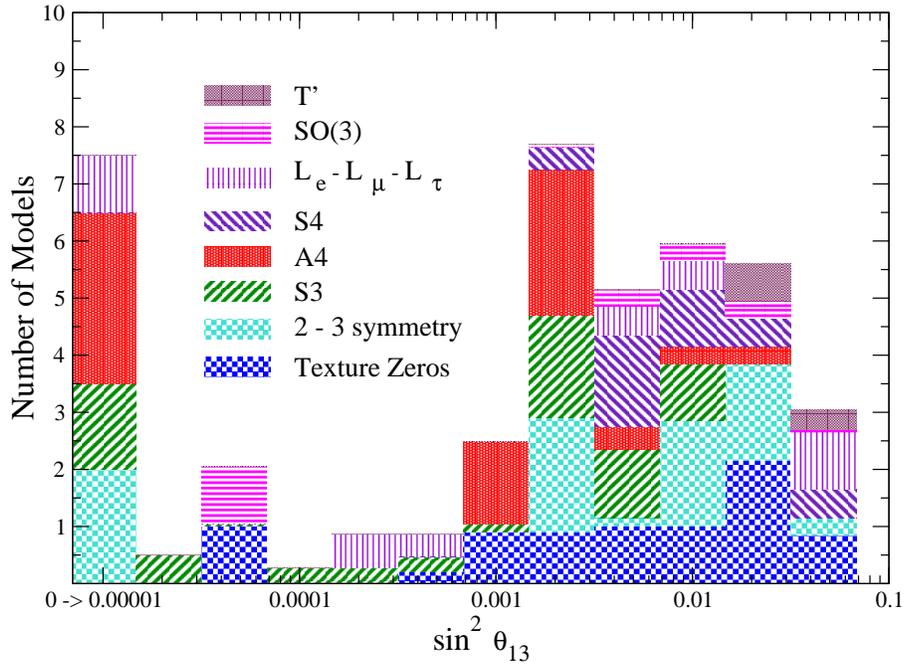}}
\caption{Predictions of $\sin^2 \theta_{13}$ for the lepton flavor
  models considered.}
\label{fig:LFmodels}
\end{center}
\end{figure}
  
\begin{figure}
\begin{center}
\mbox{\epsfig{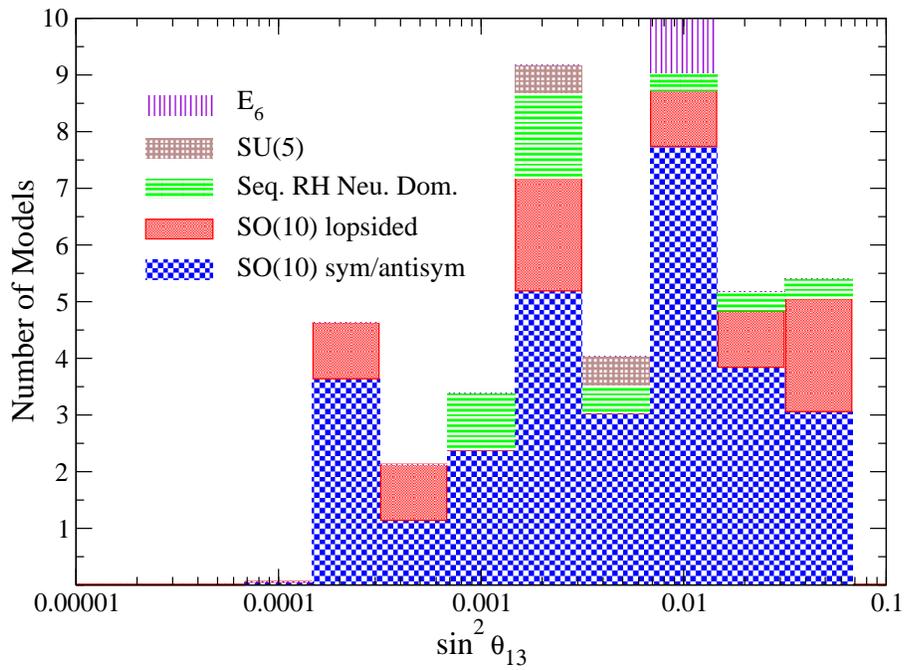}}
\caption{Predictions of $\sin^2 \theta_{13}$ for the SO(10)
  models considered.}
\label{fig:SO10models}
\end{center}
\end{figure}

\begin{figure}
\begin{center}
\mbox{\epsfig{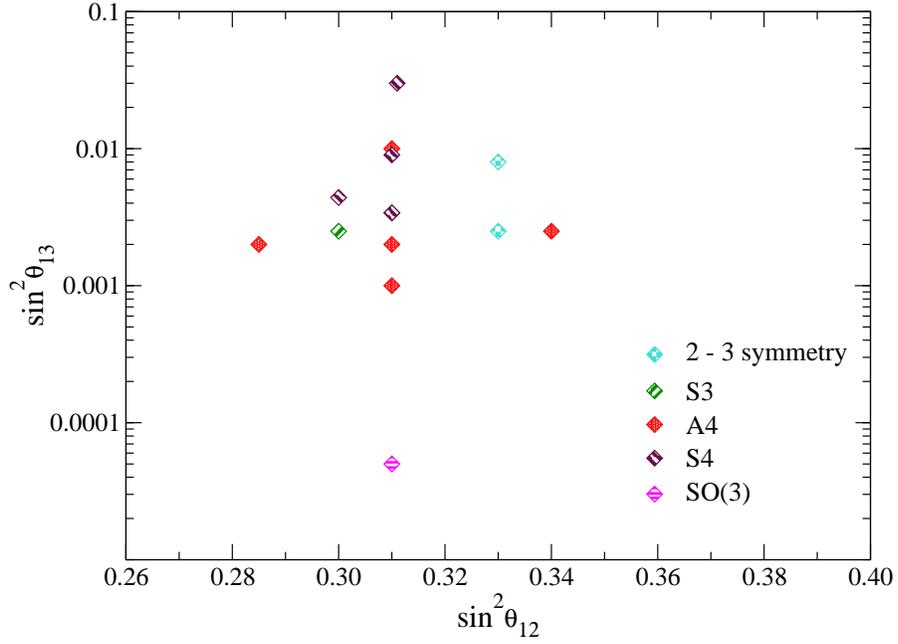}}
\caption{Predictions of the $\sin^2 \theta_{23}\ vs.\ \sin^2 \theta_{12}$
  distribution of central values for the discrete flavor symmetry 
  models considered.}
\label{fig:LFM1213}
\end{center}
\end{figure}

\begin{figure}
\begin{center}
\mbox{\epsfig{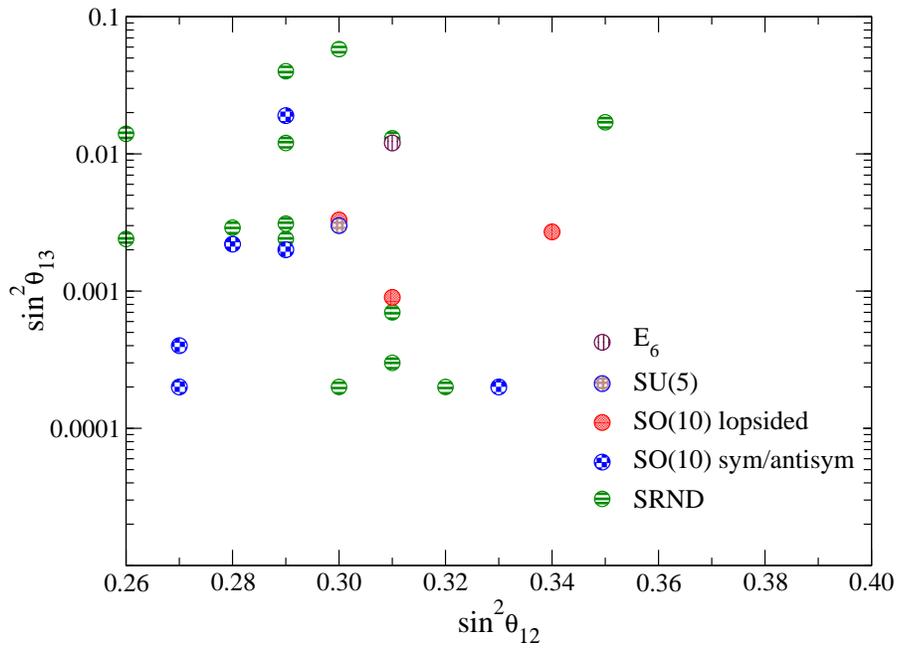}}
\caption{Predictions of the $\sin^2 \theta_{23}\ vs.\ \sin^2 \theta_{12}$
  distribution of central values for the grand unified  
  models considered.}
\label{fig:GUT1213}
\end{center}
\end{figure}

\begin{figure}
\begin{center}
\mbox{\epsfig{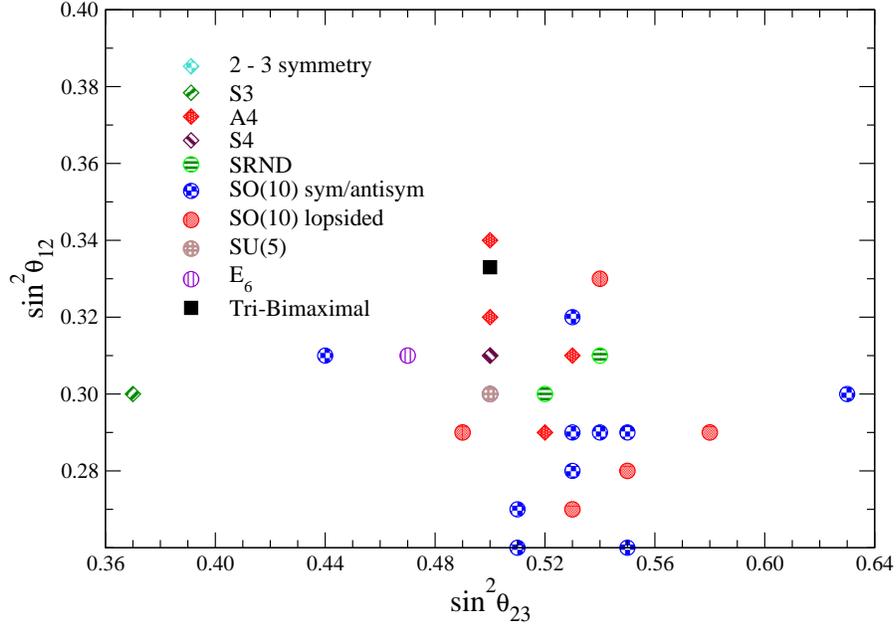}}
\caption{Predictions of the $\sin^2 \theta_{12}\ vs.\ \sin^2 \theta_{23}$
  distribution of central values for both types of   
  models considered.}
\label{fig:LFMGUT2312}
\end{center}
\end{figure}

\section{Other Tests}
\vspace*{-0.2in}
\subsection{Nature of Tri-bimaximal Mixing}
As pointed out earlier, many of the GUT models were based on
continuous and/or discrete flavor symmetries with no aim in mind to 
reproduce tri-bimaximal mixing at leading order.  This raises 
the issue whether tribimaximal mixing is a hidden symmetry 
which is softly broken or just an accidental symmetry of nature.

In order to pursue this issue, the author in collaboration with 
Werner Rodejohann adopted a model-independent approach\cite{ar1}.
In the lepton flavor basis, deviations from tri-bimaximal mixing
were considered by perturbing each element of the neutrino
mass matrix by up to 20\%:
\begin{equation}
m_\nu = \left(\begin{array}{ccc} A(1 + \epsilon_1) & B(1 + \epsilon_2) 
& B(1 + \epsilon_3)\\
   \cdot & \frac{1}{2}(A+B+D)(1 + \epsilon_4) & \frac{1}{2}(A+B-D)(1 + 
   \epsilon_5)\\
   \cdot & \cdot & \frac{1}{2}(A+B+D)(1+\epsilon_6)\\ \end{array}\right)
\label{TBMpert}
\end{equation}

\noindent  Recall that for TBM mixing,   $A = \frac{1}{3}(2m_1 + 
  m_2 e^{-2i\alpha}),
\quad B = \frac{1}{3}(m_2 e^{-2i\alpha} - m_1),\quad D = m_3 e^{-2i\beta}$.  

\begin{figure}
\begin{center}
\includegraphics*[width=12.0cm]{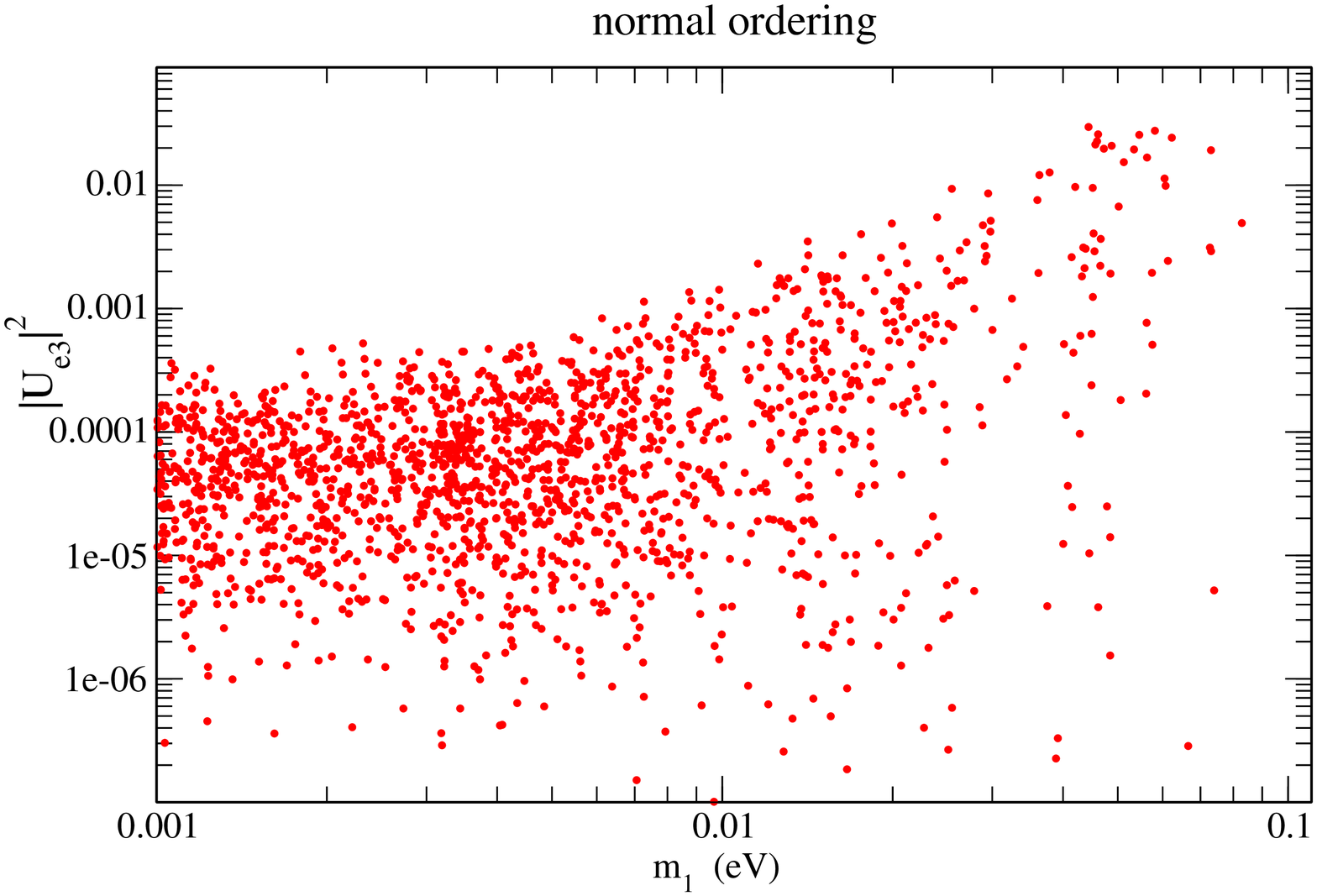}
\caption{Scatterplot for $\sin^2 \theta_{13}\ vs.\ m_1$
  distribution for normal ordering of perturbed tri-maximal mixing.}
\label{fig:brTBM_nh}
\end{center}
\end{figure}

\begin{figure}
\begin{center}
\includegraphics*[width=12.0cm]{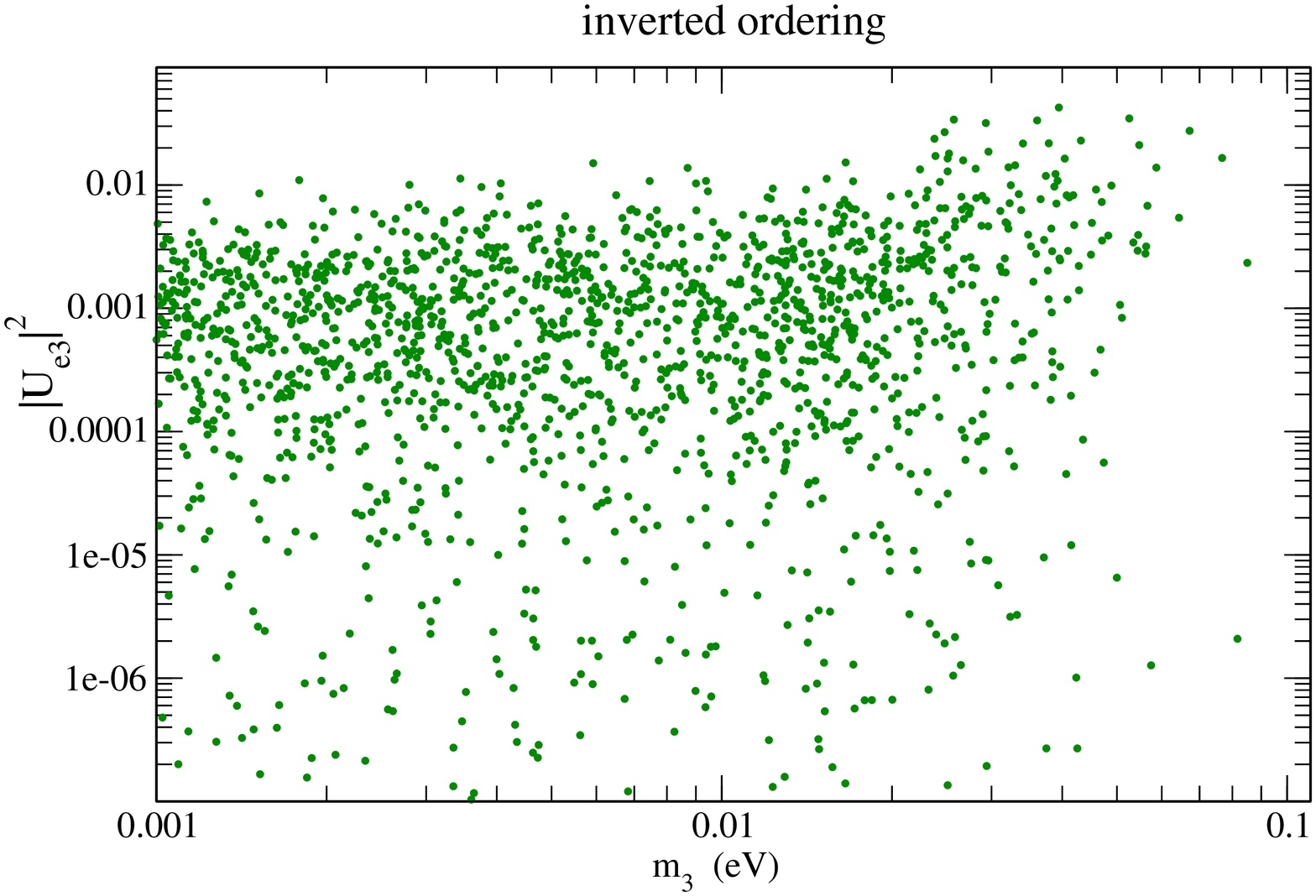}
\caption{Scatterplot for $\sin^2 \theta_{13}\ vs.\ m_1$
  distribution for inverted ordering of perturbed tri-maximal mixing.}
\label{fig:brTBM_ih}
\end{center}
\end{figure}

Scatterplots are then constructed with points chosen according to the 
following prescription:  
Start with the central best values for the mass differences in 
Eq. (\ref{eq:data}), hold $m_3$ fixed for normal hierarchy or $m_2$
fixed for inverted hierarchy, and let the other masses vary by up to
20\%; 
vary the Majorana phases in their full ranges;
and vary each $\epsilon_i$ within $|\epsilon_i| \leq 0.2$
for its full phase range. 
For each choice of parameters the resulting mass matrix is diagonalized
and, if the outcome is within the current 3$\sigma$ ranges quoted 
in Eq. (\ref{eq:data}), the point is kept.

The resulting scatterplots of $|U_{e3}|^2\ vs.\ m_1$ for normal ordering
and $vs.\ m_3$ for inverted ordering are shown in Figs. 6 and 7, 
respectively.  From Fig. 6 one sees that $|U_{e3}|^2$ remains below 
0.001 for all $m_1 < 4.5$ meV, corresponding to a normal hierarchy,
and only increases above that value once larger values of $m_1$
appear, corresponding to normal ordering until quasi-degenerate 
neutrino masses occur.  In Fig. 7 for the inverted hierarchy and 
ordering, the corresponding bound is noticeably higher at 0.01.

\begin{figure}
\begin{center}
\includegraphics*[width=12.0cm]{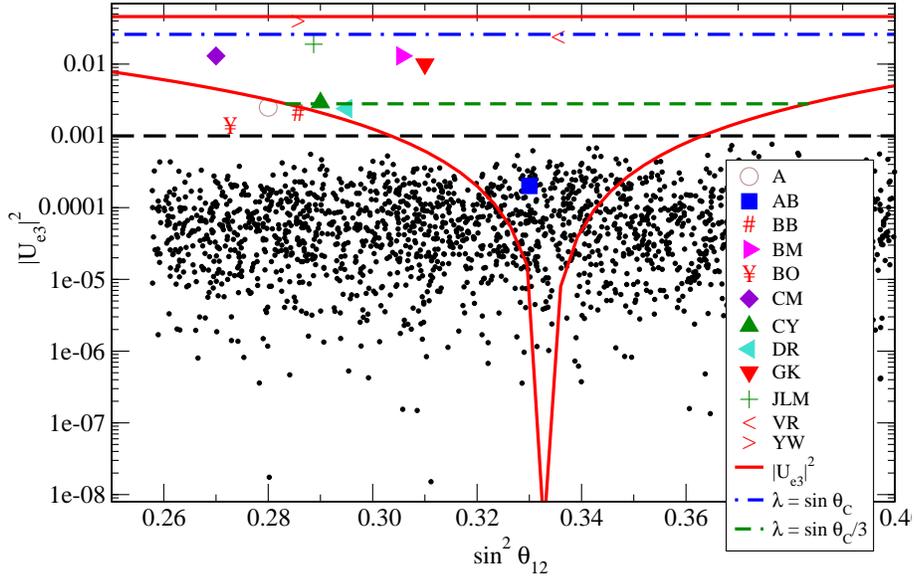}
\caption{Scatterplot for $\sin^2 \theta_{13}\ vs.\ sin^2 \theta_{12}$
  for perturbed TBM mixing and GUT model predictions.}  
\label{fig:carl_venice3}
\end{center}
\end{figure}

The scatterplot in Fig. 8  of $|U_{e3}|^2\ vs.\ \sin^2 \theta_{12}$ applies
for the normal hierarchy case with a fixed value of $m_3 = 0.050$ eV.
Again it is apparent that all points lie below 0.001.  This suggests that
for a normal hierarchy of neutrino masses, tri-bimaximal mixing is
accidental, if $\sin^2 \theta_{13}$ is found experimentally to be 
larger than the bounded deviation from zero of 0.001.  No such 
statement can be made for an inverted hierarchy, for the restricted
bound is much weaker for deviations from TBM mixing and can 
essentially extend up to nearly the present experimental limit.
Also note from Fig. 8 that no restrictions are placed on deviations
of $\sin^2 \theta_{12}$ from the TBM value of 0.333.

For comparison, we also show the results for twelve GUT models.
Note that for all but one, $\sin^2 \theta_{13}$ is projected to lie
above the softly-broken TBM mixing bound of 0.001.

However, if the charged lepton flavor matrix is rotated by one-third
the Cabibbo angle (or by the Cabibbo angle itself) from its original 
diagonal form, while the neutrino matrix keeps the TBM form, one 
finds a larger deviation of $\sin^2 \theta_{13} = 0.0029$ (0.025).
These limits are depicted by the dashed and broken lines, 
respectively, in Fig. 8.  For an arbitrary 12 rotation of the charged
lepton mass matrix from the diagonal form, the acceptable points
lie between the solid line boundaries.\\[0.2in]

\begin{figure}[h!]
\begin{center}
\includegraphics*[width=12.0cm]{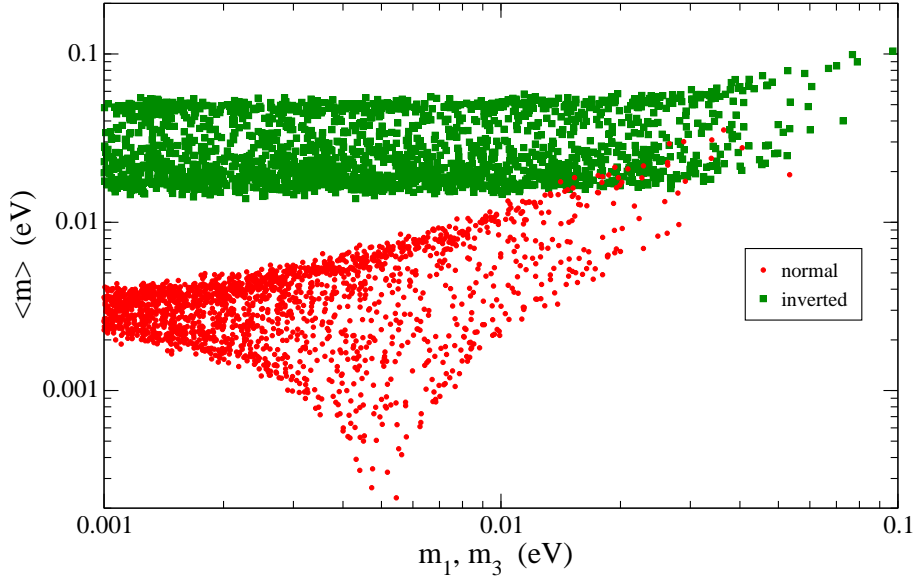}
\caption{Effective mass plot for neutrino-less double beta decay in 
  the case of perturbed tri-bimaximal mixing.}  
\label{fig:meff_brokenTBMa}
\end{center}
\end{figure}

\subsection{Neutrino-less Double Beta Decay}
Neutrino-less double beta decay provides an opportunity to test
the Majorana $vs.$ Dirac nature of the light neutrinos and 
whether the mass ordering is normal or inverted in the former 
case.  The square of the effective mass entering the decay rate
is given by 
\begin{eqnarray}
\langle m_{\beta\beta}\rangle &=& \left| \Sigma_i m_i U^2_{ei}
  \Phi^2_{ii}\right|\nonumber\\
&=& m_1 U^2_{ei} + m_2 U^2_{e2}e^{i2\alpha} + m_3 U^2_{e3} e^{i2\beta}
  \nonumber\\
&\simeq& m_1 \cos^2 \theta_{12} + m_2 \sin^2 \theta_{12} e^{i2\alpha}, 
\end{eqnarray}
where it is apparent the Majorana phases play an important role.
Since $m_1 \sim m_2 \gg m_3$ for the inverted hierarchy case, 
the $(Z,A) \rightarrow (Z+2,A) + 2 e^-$ process should occur with 
a shorter lifetime than for the normal hierarchy case.  We show 
in Fig. 9 the effective mass plot as a function of the lightest 
neutrino mass, $m_1$ ($m_3$), in the normal (inverted) ordering
case.  The plots were obtained for tri-bimaximal mixing perturbed 
as described in the previous subsection.  There is a rather clear
separation of the normal and inverted ordering distributions.

\subsection{Charged Lepton Flavor Violation}
Charged lepton flavor violation provides one more way to differentiate
neutrino mixing models, if the Dirac and right-handed Majorana neutrino mass
matrices are specified\cite{a-c2}.  Of special interest are the limits on the 
branching ratios for $\mu \rightarrow e + \gamma$ and $\mu - e$ 
conversion, for example.  The former decay branching ratio is 
presently under test by the MEG experiment\cite{MEG} which plans to lower
the present bound of $1.2 \times 10^{-11}$ to $3 - 5 \times 10^{-13}$.  
No $\mu - e$ conversion experiment is presently underway, although plans
for one exist at both J-PARC and Fermilab.

\section{Summary} 
We have made a survey of neutrino mixing models based on  
some horizontal lepton flavor symmetry and those based on GUT models 
having a vertical family symmetry and a flavor symmetry.  We have tried to 
differentiate the models on the basis of their neutrino mass hierarchy,
mixing angles, and neutrino-less double beta predictions.
Most of the models allow either mass hierarchy with the exceptions
being just normal for the type I seesaw models and only inverted
for the conserved $L_e - L_\mu - L_\tau$ models.

For both types of models our study indicates that the upcoming 
Double CHOOZ and Daya Bay reactor experiments will be able 
to eliminate roughly two-thirds of the models surveyed, if their 
planned sensitivity reaches $\sin^2 2\theta_{13} \simeq  0.001$ 
and no depletion of the $\bar{\nu}_e$ flux is observed.  However,
no smoking gun apparently exists to rule out many types of models
based on accurate data for $\sin^2 \theta_{13}$ alone, should 
evidence for a depletion be found.  Of the order of 10 - 20 models
have similar values for this mixing angle in the 0.001 - 0.05
interval.  These results for the $\sin^2 \theta_{13}$ distributions 
involve more models but are somewhat similar to those obtained in an 
earlier survey published in 2006 in collaboration with Mu-Chun 
Chen\cite{a-c}.  Only the lepton flavor models appear to lead to 
extremely small values of $\sin^2 \theta_{13} \lsim 10^{-4}$.

Most models prefer $\sin^2 \theta_{12} \lsim 0.31$ rather than
0.333 for tri-bimaximal mixing in agreement with the present 
best value of 0.312.  On the other hand, most models prefer
$\sin^2 \theta_{23} \geq 0.50$ compared with a best fit 
value of 0.466.  

Effective mass plots for perturbed tri-bimaximal mixing show 
a clear separation of the normal and inverted ordering 
distributions, so accurate neutrino-less double beta decay
experiments should be decisive.

It is clear that very accurate determination of the three mixing
angles and eventually the three CP-violating phases will be
required to pin down the most viable models.

\newpage
\section{Acknowledgements}
The author thanks Milla Baldo Ceolin for her kind invitation and support 
to participate in the XIII International Workshop on ``Neutrino Telescopes''.
It is a pleasure to acknowledge Mu-Chun Chen for her collaboration on the 
original 2006 survey and to thank Werner Rodejohann for his important 
contributions to Sect. 6.


\begin{thebibliography}{99}
\vspace*{-0.2in}

\bibitem{a-c} C. H. Albright and M.-C. Chen, {\it Phys. Rev.} {\bf D74}
  (2006) 113006.

\bibitem{fogli}  G. L. Fogli, E. Lisi, A. Marrone, A. Palazzo and 
  A. M. Rotunno, {\it Phys. Rev. Lett.} {\bf 101} (2008) 141801;
  arXiv:0809.2936 [hep-ph].

\bibitem{tbm}    P. F. Harrison, D. H. Perkins and W. G. Scott, 
  {\it Phys. Lett.} {\bf B535} (2002) 163.

\bibitem{masssum} O. Host, O. Lahav, F. B. Abdalla and K. Eitel, 
  {\it Phys. Rev.} {\bf D76} (2007) 113005.

\bibitem{PMNS}   B. Pontecorvo, {\it Zh. Eksp. Teor. Fiz.} {\bf 33} (1957)
  549 [{\it Sov. Phys. JETP} {\bf 6} (1957) 429]; Z. Maki, M. Nakagawa
  and S. Sakata, {\it Prog. Theor. Phys.} {\bf 28} (1962) 870.

\bibitem{pdb} C. Amsler {\it et al.}, [Particle Data Group], {\it Phys. Lett.} 
  {\bf B667} (2008) 1.

\bibitem{s-k}  K. S. Hirata {\it et al.}, [Kamiokande-II Collab.],
  {\it Phys. Lett.} {\bf B280} (1992) 146.

\bibitem{textzero} P. Ramond, R. G. Roberts and G. G. Ross, {\it Nucl. Phys.}
  {\bf B406} (1993) 19.

\bibitem{e-mu-tau} R. N. Mohapatra, {\it J. High Energy Phys.} {\bf 10} (2004)
  027.

\bibitem{f-n}  C. D. Froggatt and H. B. Nielsen, {\it Nucl. Phys.} {\bf B147}
  (1979) 277.

\bibitem{lam}  C. S. Lam, {\it Phys. Rev.} {\bf D78} (2008) 073015.

\bibitem{minimal}  K. S. Babu and R. N. Mohapatra, {\it Phys. Rev. Lett.}
  {\bf 70} (1993) 2845.

\bibitem{lopsided} K. S. Babu and S. M. Barr, {\it Phys. Lett.} {\bf B381} 
  (1996) 202; C. H. Albright, K. S. Babu and S. M. Barr, {\it Phys. Rev. 
  Lett.} {\bf 81} (1998) 1167.

\bibitem{alb}  C. H. Albright, {\it Phys. Lett.} {\bf B599} (2004) 285.

\bibitem{famflav}  M.-C. Chen and K. T. Mahanthappa, {\it Phys. Lett.} 
  {\bf B652} (2007) 34; G. Altarelli, F. Feruglio and C. Hagedorn, {\it JHEP}
  {\bf 0803} (2008) 052; F. Bazzocchi, M. Frigerio and S. Morisi, 
  {\it Phys. Rev.} {\bf D78} (2008) 116018.

\bibitem{GL1} W. Grimus and L. Lavoura, {\it J. Phys.} {\bf G31} (2005) 693. 

\bibitem{WY}  A. Watanabe and K. Yoshioka, {\it JHEP} {\bf 0605} (2006) 044.

\bibitem{CPP} B. C. Chauhan, J. Pulido and M. Picariello, {\it Phys. Rev.}
  {\bf D73} (2006) 053003.

\bibitem{BM}  K. S. Babu and R. N. Mohapatra, {\it Phys. Lett.} {\bf B532}
  (2002) 77.

\bibitem{GMN1} H. S. Goh, R. N. Mohapatra and S.-P. Ng, {\it Phys. Lett.}
  {\bf B542} (2002) 116.

\bibitem{PR}  S. T. Petcov and W. Rodejohann, {\it Phys. Rev.} {\bf D71}
  (2005) 073002.

\bibitem{GL2} W. Grimus and L. Lavoura, {\it J. Phys.} {\bf G31} (2005) 683.

\bibitem{RS}  W. Rodejohann and M. A. Schmidt, {\it Phys. Atom. Nucl.} 
  {\bf 69} (2006) 1833.

\bibitem{MN}  K. Matsuda and H. Nishiura, {\it Phys. Rev.} {\bf D73} (2006)
  013008.

\bibitem{AKKL} Y. H. Ahn, S. K. Kang, C. S. Kim and J. Lee, {\it Phys. Rev.}
  {\bf D73} (2006) 093005.

\bibitem{SRB} N. N. Singh, M. Rajkhowa and A. Borah, {\it J. Phys.} {\bf G35}
  (2007) 345.

\bibitem{BY}  T. Baba and M. Yasue, {\it Phys. Rev.} {\bf D77} (2008) 075008.

\bibitem{KMMR-J} J. Kubo, A. Mondragon, M. Mondragon and E. Rodriguez-Jauregui,
  {\it Prog. Theor. Phys.} {\bf 109} (2003) 795.

\bibitem{CFM} S.-L. Chen, M. Frigerio and E. Ma, {\it Phys. Rev.} {\bf D70}
  (2004) 073008.

\bibitem{T}   T. Teshima, {\it Phys. Rev.} {\bf D73} (2006) 045019.

\bibitem{TY}  M. Tanimoto and T. Yanagida, {\it Phys. Lett.} {\bf B633} 567.

\bibitem{MNY} R. N. Mohapatra, S. Nasri and H.-B. Yu, {\it Phys. Lett.} 
  {\bf B639} (2006) 318.

\bibitem{MMP} A. Mondragon, M. Mondragon and E. Peinado, {\it AIP Conf. Proc.}
  {\bf 1026} (2008) 164.

\bibitem{MC}  M. Mitra and S. Choubey, {\it Phys. Rev.} {\bf D78} (2008) 
  115014.

\bibitem{Ma1}   E. Ma, {\it Mod. Phys. Lett.} {\bf A20} (2005) 2601.

\bibitem{ABGMP} B. Adhikary, B. Brahmachari, A. Ghosal, E. Ma and M. K. Parida,
  {\it Phys. Lett.} {\bf B638} (2006) 345.

\bibitem{AG1}  B. Adhikary and A. Ghosal, {\it Phys. Rev.} {\bf D75} (2007)
  073020.

\bibitem{HT}  M. Honda and M. Tanimoto, {\it Prog. Theor. Phys.} {\bf 119}
  (2008) 583.

\bibitem{AG2} B. Adhikary and A. Ghosal, {\it Phys. Rev.} {\bf D78} (2008)
  073007.

\bibitem{L}   Y. Lin, arXiv:0804.2867 [hep-ph].

\bibitem{Ma2}  E. Ma, {\it Phys. Lett.} {\bf B671} (2009) 366.

\bibitem{MPR} R. N. Mohapatra, M. K. Parida and G. Rajasekaran, 
  {\it Phys. Rev.} {\bf D69} (2004) 053007.

\bibitem{HLM} C. Hagedorn, M. Lindner and R. N. Mohapatra, {\it JHEP}
  {\bf 0606} (2006) 042.

\bibitem{Z}   H. Zhang, {\it Phys. Lett.} {\bf B655} (2007) 132.

\bibitem{M}  I. Masina, {\it Phys. Lett.} {\bf B633} (2006) 134.

\bibitem{W}   Y.-L. Wu, {\it Phys. Rev.} {\bf D77} (2008) 113009.

\bibitem{FM}  P. H. Frampton and S. Matsuzaki, arXiv:0902.1140 [hep-ph].

\bibitem{D}   R. Dermisek, {\it Phys. Rev.} {\bf D70} (2004) 033007.

\bibitem{K}   S. F. King, {\it JHEP} {\bf 0508} (2005) 105.

\bibitem{H}   N. Haba, {\it JHEP} {\bf 0605} (2006) 030.

\bibitem{EH}  M.-T. Eisele and N. Haba, {\it Phys. Rev.} {\bf D74} (2006)
  073007.

\bibitem{BRT} T. Blazek, S. Raby and K. Tobe, {\it Phys. Rev.} {\bf D62} (2000) 
  055001.

\bibitem{BW}  W. Buchmuller and D. Wyler, {\it Phys. Lett.} {\bf B521} (2001) 
  291.

\bibitem{SP}  N. N. Singh and M. Patgiri, {\it Int. J. Mod. Phys.} {\bf A17} 
  (2002) 3629.

\bibitem{Ra}  S. Raby, {\it Phys. Lett.} {\bf B561} (2003) 119.

\bibitem{BO}  M. Bando and M. Obara, {\it Prog. Theor. Phys.} {\bf 109} (2003) 
  995.

\bibitem{O}   N. Oshimo, {\it Nucl. Phys.} {\bf B668} (2003) 258.
 
\bibitem{KR}  S. F. King and G. G. Ross, {\it Phys. Lett.} {\bf B574} (2003) 
  239.

\bibitem{Ro}  W. Rodejohann, {\it Phys. Lett.} {\bf B579} (2004) 127.

\bibitem{GMN2} H. S. Goh, R. N. Mohapatra and S.-P. Ng, {\it Phys. Rev.} 
  {\bf D68} (2003) 115008.

\bibitem{YW}  W.-M. Yang and Z.-G. Wang, {\it Nucl. Phys.} {\bf B707} (2005) 
  87.

\bibitem{CM1} M.-C. Chen and K. T. Mahanthappa, arXiv:hep-ph/0409165.

\bibitem{BeM} S. Bertolini and M. Malinsky, {\it Phys. Rev.} {\bf D72} (2005) 
  055021.

\bibitem{BaM} K. S. Babu and C. Macesanu, {\it Phys. Rev.} {\bf D72} (2005) 
  115003.

\bibitem{DR} R. Dermisek and S. Raby, {\it Phys. Lett.} {\bf B622} (2005) 327.

\bibitem{VR}  I. M. Varzielas and G. G. Ross, {\it Nucl. Phys.} {\bf B733} 
  (2006) 31.

\bibitem{DMM} B. Dutta, Y. Mirmura and R. N. Mohapatra, {\it Phys. Rev.} 
  {\bf D72} (2005) 075009.

\bibitem{ShT}  Q. Shafi and Z. Tavartkiladze, {\it Phys. Lett.} {\bf B633} 
  (2006) 595.

\bibitem{BN}  Z. Berezhiani and F. Nesti, {\it JHEP} {\bf 0603} (2006) 041.

\bibitem{BMSV} B. Bajc, A. Melfo, G. Senjanovic and F. Vissani, {\it Phys. 
  Lett.} {\bf B634} (2006) 272.

\bibitem{DHR} R. Dermisek, M. Harada and S. Raby, {\it Phys. Rev.} {\bf D74} 
  (2006) 035011.

\bibitem{KM}  S. F. King and M. Malinsky, {\it JHEP} {\bf 0611} (2006) 071.

\bibitem{CY}  Y. Cai and H.-B. Yu, {\it Phys. Rev.} {\bf D74} (2006) 115005.

\bibitem{GK1} W. Grimus and H. Kuhbock, {\it Eur. Phys. J.} {\bf C51} (2007) 
  721.

\bibitem{FMN} T. Fukuyama, K. Matsuda and H. Nishiura, {\it Int. J. Mod. 
  Phys.} {\bf A22} (2007) 5325.

\bibitem{GK2} W. Grimus and H. Kuhbock, {\it Phys. Rev.} {\bf D77} (2008) 
  055008.

\bibitem{Mo}  R. G. Moorhouse, {\it Phys. Rev.} {\bf D77} (2008) 053006.

\bibitem{P}   M. K. Parida, {\it Phys. Rev.} {\bf D78} (2008) 053004.

\bibitem{BR} P. D. Bari and A. Riotto, {\it Phys. Lett.} {\bf B671} (2009) 
  462.

\bibitem{Mae}  N. Maekawa, {\it Prog. Theor. Phys.} {\bf 106} (2001) 401.

\bibitem{AB}  C. H. Albright and S. M. Barr, {\it Phys. Rev.} {\bf D64} 
  (2001) 073010.

\bibitem{BB}  K. S. Babu and S. M. Barr, {\it Phys. Lett.} {\bf B525} (2002) 
  289.

\bibitem{A}   C. H. Albright, {\it Phys. Rev.} {\bf D72} (2005) 013001; 
  {\it erratum-ibid.} {\bf D74} (2006) 039903.

\bibitem{JLM} X. Ji, Y. Li and R. N. Mohapatra, {\it Phys. Lett.} {\bf B633} 
  (2006) 755.

\bibitem{CM2} M.-C. Chen and K. T. Mahanthappa, {\it Phys. Lett.} {\bf B652} 
  (2007) 34.

\bibitem{StT} B. Stech and Z. Tavartkiladze, {\it Phys. Rev.} {\bf D77}
  (2008) 076009.

\bibitem{FHLR} M. Frigerio, P. Hosteins, S. Lavignac and A. Romanino, 
  {\it Nucl. Phys.} {\bf B806} (2009) 84.

\bibitem{HSS} C. Hagedorn, M. A. Schmidt and A. Yu. Smirnov, {\it Phys. Rev.} 
  {\bf D79} (2009) 036002.

\bibitem{reac} C. E. Lane, [Double CHOOZ Collab.], in {\it Proceedings of 
   the 34th International Conference on High Energy Physics} (Philadelphia, 
   2008); M.-C. Chu [Daya Bay Collab.], {\it ibid.}

\bibitem{T2KNOVA} I. Kato, [T2K Collab.], {\it J. Phys.: Conf. S.} {\bf 136}
  (2008) 022018; R. Ray, [NO$\nu$A Collab.], {\it J. Phys.: Conf. S.}
  {\bf 136} (2008) 022019.

\bibitem{ar1} C. H. Albright and W. Rodejohann, {\it Phys. Lett.} {\bf B665}
  (2008) 378.

\bibitem{a-c2} C. H. Albright and Mu-Chun Chen, {\it Phys. Rev.} {\bf D77}
  (2008) 113010.

\bibitem{MEG} T. Mori, {\it ECONFC} {\bf 060409} (2006) 034. 

\end{thebibliography}
\end{document}